\documentclass[conference,a4paper]{IEEEtran}

\usepackage{amsthm}
\usepackage{amssymb}
\usepackage[utf8]{inputenc}
\usepackage{amsmath}
\usepackage{graphicx}
\usepackage{epsfig}
\usepackage[tight,footnotesize]{subfigure}
\usepackage{url}
\usepackage{tikz}
\usepackage{algorithm2e}

\makeatletter
\def\ps@headings{%
\def\@oddhead{\mbox{}\scriptsize\rightmark \hfil \thepage}%
\def\@evenhead{\scriptsize\thepage \hfil \leftmark\mbox{}}%
\def\@oddfoot{}%
\def\@evenfoot{}}
\makeatother

\usepackage{cite}

\ifCLASSINFOpdf
\else
\fi
\usepackage{url}

\IEEEoverridecommandlockouts

\begin{document}
%
\title{Compressive Data Aggregation on Mobile Wireless Sensor Networks for Sensing in Bike Races}
\author{
\IEEEauthorblockN{Wei Du\thanks{This work is funded by the French government under the grant FUI-SMACS.}, Jean-Marie Gorce, Tanguy Risset}
\IEEEauthorblockA{Univ Lyon, INSA Lyon, Inria, CITI, F-69621 Villeurbanne, France\\Email: \{wei.du,jean-marie.gorce,tanguy.risset\}@insa-lyon.fr} \and
\IEEEauthorblockN{Matthieu Lauzier, Antoine Fraboulet}
\IEEEauthorblockA{HIKOB, France\\Email: \{matthieu.lauzier,antoine.fraboulet\}@hikob.com}
}

\maketitle

\begin{abstract}
This paper presents an efficient approach for collecting data in mobile wireless sensor networks which is specifically designed to gather real-time information of bikers in a bike race. The approach employs the recent HIKOB sensors for tracking the GPS position of each bike and the problem herein addressed is to transmit this information to a collector for visualization or other processing. Our approach exploits the inherent correlation between biker motions and aggregates GPS data at sensors using compressive sensing (CS) techniques. We enforce, instead of the standard signal sparsity, a spatial sparsity prior on biker motion because of the grouping behavior ({\em peloton}) in bike races. The spatial sparsity is modeled by a graphical model and the CS-based data aggregation problem is solved using linear programming. Our approach, integrated in a multi-round opportunistic routing protocol, is validated on data generated by a bike race simulator using trajectories of motorbikes obtained from a real race, the Paris-Tours 2013.
\end{abstract}

\section{Introduction}
Modern sensing technologies already enable real-time monitoring of various environments or human activities in applications such as traffic control, health care, smart homing and sports. For example, GPS sensors have been used in recent bike races including the Tour de France for collecting biker locations, and such data can be exploited to enhance live TV broadcasting or to analyze biker performance \cite{tourdefrance_gps}.

\begin{figure}[t]
    \begin{center}
        \includegraphics[width=0.4\columnwidth]{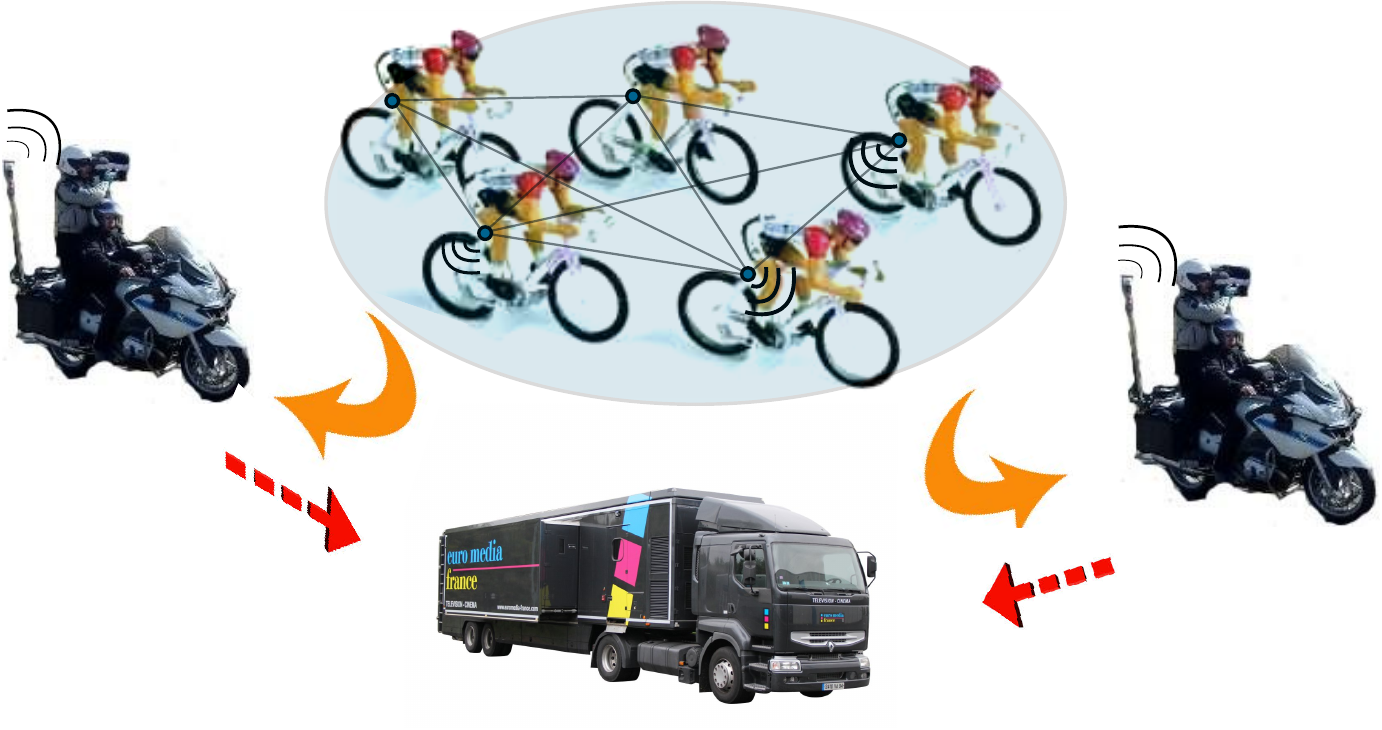}
    \end{center}
    \caption{The HIKOB infrastructure for data collection in bike races.}
    \label{fig:infra}
\end{figure}

In most applications of wireless sensor networks (WSNs), the efficiency of data collection is critical as sensors come with limited power while the radio communication is energy consuming. The context of bike racing adds additional constraints due to the high mobility of the bikers. Obviously, it is not possible to ensure a direct connection between each sensor and a common sink due e.g. to body shadowing and biker spreading. Therefore, routing data between sensors is necessary to ensure reliable data collection. However, data routing is complicated because the routes are time-varying and cannot be pre-defined. Additionally, the application requires a high refresh rate on the location measurement, typically about one reading per second in a race. 
HIKOB \cite{hikob} provides GPS sensors that meet these requirements and designed an infrastructure for data collection in bike races, illustrated in Figure \ref{fig:infra}, in which GPS data are relayed from sensors on bikes, via intermediate sinks on motorbikes surrounding the race, to the central sink on a data truck for further processing. The particular challenge we address in this paper is how to route data between sensors on bikes in an energy-efficient way.

A naive but direct solution would be a multihop opportunistic routing protocol in which each sensor transmits in several rounds not only its own data but also the data received from its neighbors. Due however to the bandwidth constraint, each sensor can only select and transmit a subset of the received data. The selection of this subset is strategic to maximize the chance of receiving all data at the sink. Since it is impossible to fully coordinate the data selection process at different sensors, it is inevitable to have redundant and missing data, i.e. some data are received for multiple times, while others are not received at all. The problem is more severe with poor radio connections which result in a high packet loss rate.

To improve the efficiency of the data collection, this paper proposes to aggregate data at sensors instead of selecting a few data, based on compressive sensing (CS) techniques \cite{cs,csintroduction}. CS is a class of methods that deal with sparse signals and has been widely used for data collection on WSNs \cite{compressivedatacollection,dataaggregation,cscluster}. However, the general CS framework in previous work is designed for static WSNs with predefined routing paths and cannot deal with packet losses which occur frequently in mobile WSNs. In contrast, we design a novel CS approach for the application of sensing in bike races. Instead of the standard signal sparsity, we enforce a problem-specific sparsity prior on the motion of the bikers. The assumption is that many bikers follow the constant velocity model and show strong group behavior. Indeed, cycling is a team sport in which bikers group and draft for the aerodynamic benefit and they accelerate only occasionally to break away from a group. Thus, we incorporate the prior information that nearby bikers tend to make synchronized moves and thus keep their relative locations in the group. This spatial sparsity is modeled by a graphical model which captures the spatial relationships between bikers. We formulate and solve the CS problem with the spatial sparsity using linear programming (LP). Furthermore, a simple multi-round routing protocol based on broadcasting and aggregation is designed and integrates seamlessly the CS method. Our approach is evaluated on data generated by a bike race simulator based on real trajectories of motorbikes obtained in the race of Paris-Tours 2013, and is shown to be able to recover accurately all data even in situations when many sensors are poorly connected.

Our technical contributions include the incorporation of a spatial prior on biker motion and the resolution of the CS problem by LP. Our approach falls in the category of modal-based CS \cite{mbcs} and CS with structured sparsity \cite{structuredsparsity} which enforces additional structures or priors on the sparsity. Our spatial prior is a local constraint that enforces similarities on the motion of nearby bikers which is specific to our problem. In addition, we study the integration of the CS method with our multi-round opportunistic routing protocol. The rest of the paper is organized as follows. Section \ref{sec:csforwsn} introduces the idea of CS and its applications on WSNs. Section \ref{sec:csforbikeraces} describes our CS method with the spatial prior and the routing protocol. Section \ref{sec:experiments} illustrates some results on simulated data. Conclusions and future work are given in section \ref{sec:conclusions}.

\section{CS for WSNs}
\label{sec:csforwsn}
\subsection{General CS and its applications on WSNs}
The general idea of CS is as follows \cite{cs,csintroduction}. Let $X$ be a signal of size $n$, $X=[x_1,\ldots,x_n]^T$, and $X$ is $K$-sparse in the space $\phi$, i.e. $\phi X$ contains at most $K$ significant coefficients. $\phi$ is of size $n\times n$ and degenerates to the identity matrix if $X$ is sparse itself. Let $Y$ be a measurement vector of size $k$, $Y=[y_1,\ldots,y_k]^T$, $k\ll n$, and $Y$ is formed by the projection of the $K$-sparse coefficients $\phi X$ on the random matrix $A$ of size $k\times n$, i.e. $Y=A\phi X$. The entries of $A$ are drawn i.i.d. from either Gaussian or Bernoulli distribution. The CS theory asserts that one can recover $X$ from $Y$, if $k\geqslant O(Klog(n/K)$, by solving for the noiseless case
 \begin{equation}
 \min ||\phi X||_1 \qquad s.t.\ Y=A\phi X,
 \label{eq:cs_noiseless}
 \end{equation}
or for the noisy case
 \begin{equation}
 \min ||\phi X||_1 \qquad s.t.\ ||Y-A\phi X||_2^2\leqslant\epsilon.
 \label{eq:cs_noisy}
 \end{equation}
The above optimization can be solved easily by e.g. lasso \cite{lasso}.

\begin{figure}[t]
    \begin{center}
        \includegraphics[width=0.6\columnwidth]{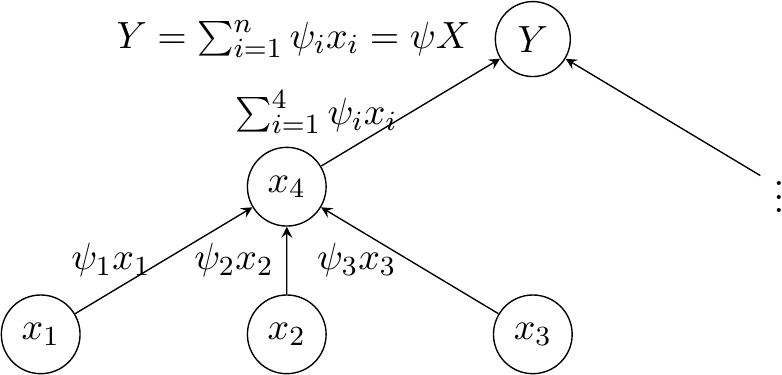}
    \end{center}
    \caption{CS for data collection on a routing tree of a WSN. Define $\psi=A\phi$ and its $i$th column $\psi_i$ is stored at sensor $i$, $\psi=[\psi_1,\ldots,\psi_n]$. Sensor $i$ transmits a vector of size $k$ which is the combination of its own data $\psi_i x_i$ and of all data from the offspring, and the sink receives $Y=\psi X=A\phi X$.}
    \label{fig:csroutingtree}
\end{figure}

For applications of WSNs, CS is promising as it can reduce the number of data transmission from $n$ to $k$. Previous work on CS for WSNs \cite{compressivedatacollection,dataaggregation,cscluster} employed a similar framework, illustrated in Figure \ref{fig:csroutingtree}, in which the compressive data collection is done on a routing tree, i.e. each sensor aggregates in one packet its own data and all data received from the children and transmits the aggregated data to its parent.

There are two issues with the framework in Figure \ref{fig:csroutingtree}. First, it works only for static WSNs where the routing path for each sensor is fixed and known by the sink. For our application, due to the high mobility of the bikers, we can only perform CS with no predefined routing. Second, it does not handle packet losses due e.g. to poor radio connections which are more severe for mobile WSNs. Packet losses destroy the sparsity pattern of the data since the lost data can only be treated as 0, illustrated in Figure \ref{fig:cs_coefficients}. Only when $X$ is directly sparse, i.e. $\phi$ is an identity matrix, packet losses do not cause big problems, which is barely true for most types of data including the GPS data in our application.

\subsection{CS with Packet Losses}
\begin{figure}[t]
\centering
\subfigure[Impact of packet losses to the sparsity pattern.]{%
        \includegraphics[height=2cm, width=\columnwidth]{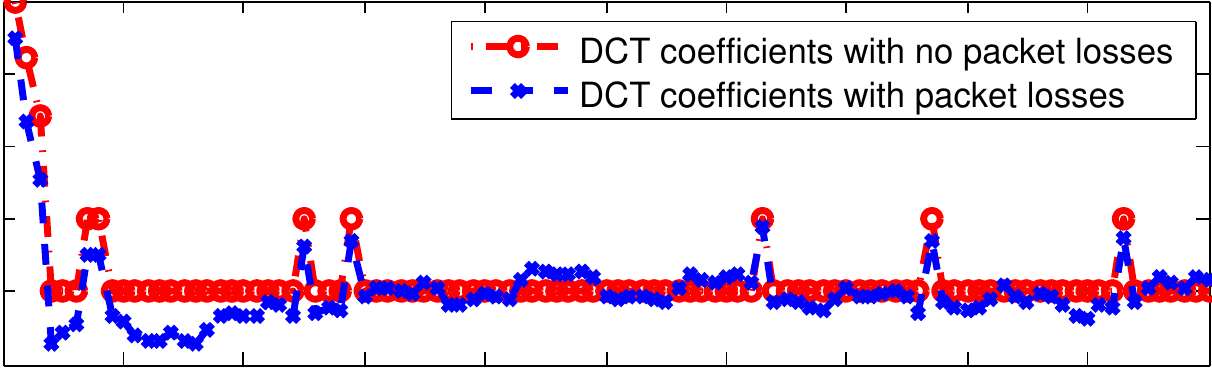}
\label{fig:cs_coefficients}}
\subfigure[Comparison of the reconstruction accuracy.]{%
        \includegraphics[height=2cm, width=\columnwidth]{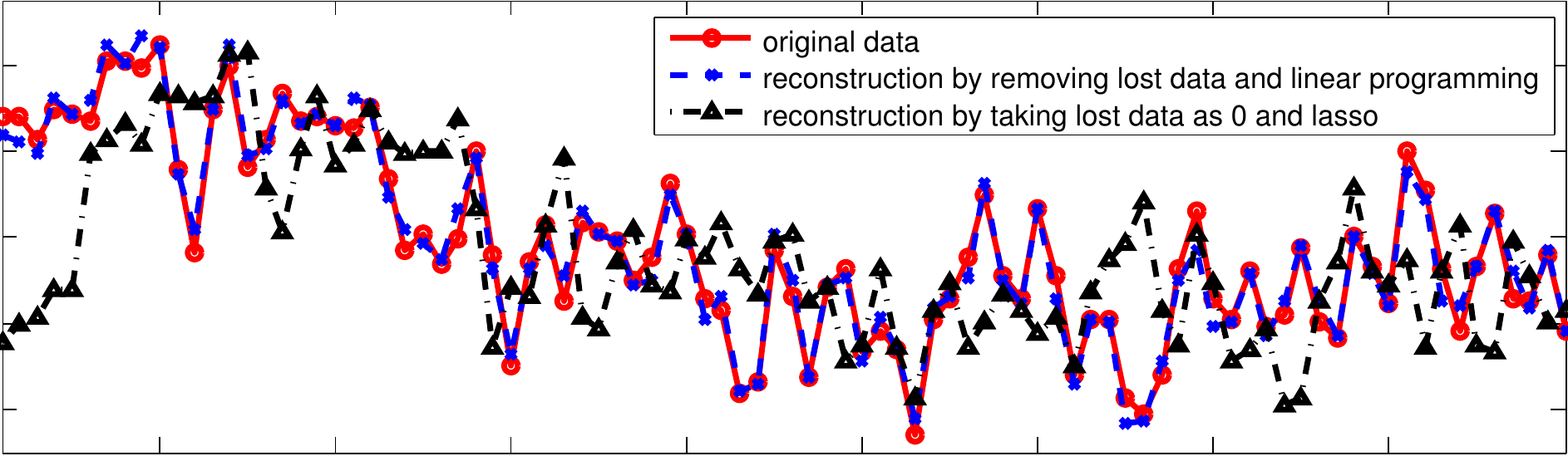}
\label{fig:cs_reconstruction}}
\caption{We generate a signal which is 10-sparse in the discrete cosine transform (DCT) domain, shown as the red line in (a). We simulate 10 packet losses by setting randomly 10 data in the signal to 0, and the DCT coefficients become non-sparse, shown as the blue line in (a). In practice, packet losses close to the root of the routing tree cause much more severe problems. We compare the accuracy of the reconstruction with packet losses. It can be seen that taking lost data as 0 doesn't work well, shown as the black line in (b), due to the non-sparse pattern in (a), whereas linear programming with removing lost data produces a fairly good reconstruction, shown as the blue line in (b).}
\label{fig:cs_packetloss}
\end{figure}

We first provide a general CS method which handles packet losses naturally. Instead of aggregating $\phi X$, we aggregate $X$ directly and randomly without doing the transform $\phi$ while assuming the sparsity of $\phi X$ , i.e.
 \begin{equation}
 \min ||\phi X||_1 \qquad s.t.\ Y=A X.
 \label{eq:cs_packetloss}
 \end{equation}
The problem of packet losses is introduced by the fact that $\phi$ is predefined for transforming $X$ which is of size $n$. With packet losses, we are in fact dealing with a signal of smaller size. Let $X^{\prime}$ be the vector of all received $x_i$s. Clearly, eq. \ref{eq:cs_packetloss} can deal with packet losses because the sink receives $Y^{\prime}=A^{\prime}X^{\prime}$, $A^{\prime}$ being the remaining of $A$ with the columns corresponding to the lost $x_i$ removed, and finds the right $\phi^{\prime}$ for transforming $X^{\prime}$. Eq. \ref{eq:cs_packetloss} cannot be solved by standard lasso, but can be converted to a problem of linear programming (LP) as follows,
 \begin{eqnarray}
 \min\ \  & \boldsymbol{1}^T \delta. \\
 s.t.\ \  & Y=A X \nonumber \\
 & -\delta \leqslant \phi X \leqslant \delta \nonumber\\
 & \delta \geqslant 0\nonumber
 \label{eq:cs_packetloss_lp}
 \end{eqnarray}
Note that $\delta$ is a vector of length $n$ which defines the bound on each entry of $|\phi X|$. A comparison of the reconstruction results is shown in Figure \ref{fig:cs_reconstruction}, which illustrates the better performance of eq. \ref{eq:cs_packetloss}. While $X$ is not directly sparse, eq. \ref{eq:cs_packetloss} is intuitively solvable as eq. \ref{eq:cs_noiseless} and \ref{eq:cs_noisy}, because it is equivalent to the following formulation,
 \begin{equation}
 \min ||C||_1 \qquad s.t.\ Y=A \phi^{-1}C,\ C=\phi X,
 \label{eq:cs_packetloss_C}
 \end{equation}
where $C$ is the sparse coefficients of $X$ in the space $\phi$.

\section{CS for Sensing in Bike Races}
\label{sec:csforbikeraces}
The keys to the design of a proper CS method for an application of WSNs include the integration of CS with the routing protocol and the choice of the sparse space $\phi$ which is generally specific to the application. This section describes our efforts to solve these issues.

\subsection{Sensor Data}

The data of interest is GPS readings from each sensor which are in the form of (longitude, latitude, altitude). In this paper, we convert each GPS reading into a quantity of relative location with respect to a starting point using the road map from which we further compute the motion of the bikers over time. Thus, we aim to collect from each sensor the motion data instead of the location data. In particular, for a race of $n$ bikers, let $x_i(t)$ be the instant velocity of biker $i$ at time $t$, $i=1,\ldots,n$, and $x_i(t)$ can be directly computed at each sensor from its GPS readings. Without causing confusions, we drop the time index for simplifying the notations as we consider independent data collection at each time. Note that our CS method can easily include the sparsity in the temporal domain, with little modification.

\subsection{Spatial Sparsity}

To perform CS, we need to know in which space $\phi$ the data of interest is sparse. For our data of biker motion, we assume that the bikers in a race follow the constant velocity model and show strong group behavior in order to save energy. Thus, we define a spatial sparsity which enforces synchronized motion between nearby bikers. The motivation comes from the flocking behavior which, while looking complicated, is controlled by three simple rules \cite{flocking}:
\begin{itemize}
 \item Separation: avoid crowding neighbors.
 \item Alignment: steer towards average heading of neighbors.
 \item Cohesion: steer towards average position of neighbors.
\end{itemize}
The grouping behavior of bikers resembles in some ways the flocking behavior of birds both of which aim at the aerodynamic benefit.

To model the spatial sparsity, we first construct a graph modeling the spatial relationships between the bikers. Here, we simply connect each biker with her/his k-nearest neighbors using their positions in the previous time instant. Then, we penalize the differences between the velocities of neighboring bikers under the constraint of $Y=AX$. That is, our CS formulation based on the spatial sparsity is defined as
 \begin{equation}
 \min \sum_{ij\in E,i<j} ||x_i-x_j||_1 \qquad s.t.\ Y=A X,
 \label{eq:cs_spatialsparsity}
 \end{equation}
where $E$ is the set of edges in the k-nearest neighbor graph. The spatial sparsity is specific to our application and related to the graph Laplacian regularization for graph signal processing \cite{graphsignal}. To see this, a Laplacian of a graph is defined as a matrix $L$ with entries of
\begin{equation}
 L_{ij}=\begin{cases}
  \text{degree}(i) & i=j\\
  -1 & ij\in E,i\neq j\\
  0 & otherwise
 \end{cases}.
 \label{eq:Laplacian}
\end{equation}
Thus, the spatial sparsity $\sum_{ij\in E,i<j} ||x_i-x_j||_1$ is in nature similar to the graph Laplacian regularization $||LX||_1$ where $L$ corresponds to the sparse space $\phi$. The difference is that $||LX||_1=\sum_i ||\sum_{ij\in E,\forall j} (x_i-x_j)||_1$ enforces the total difference between each biker and the neighbors to be zero or as small as possible, whereas $\sum_{ij\in E,i<j} ||x_i-x_j||_1$ enforces the similarity between each pair of neighboring bikers which is a stronger prior on $X$.

Similarly, eq. \ref{eq:cs_spatialsparsity} can be converted to a standard LP problem, given by
 \begin{eqnarray}
 \min\ \  & \sum_{ij\in E, i<j} \delta_{ij}, \\
 s.t.\ \  & Y=A X \nonumber\\
 & -\delta_{ij} \leqslant x_i-x_j \leqslant \delta_{ij}, \quad \forall ij\in E,\ i<j \nonumber \\
 & \delta_{ij} \geqslant 0, \quad \forall ij\in E,\ i<j\nonumber
 \label{eq:cs_spatialsparsity_lp}
 \end{eqnarray}
which can be solved easily by any standard LP solver.

\subsection{Integration of CS with Routing}
 
Based on the CS formulation in eq. \ref{eq:cs_spatialsparsity}, we design a simple routing algorithm for collecting data in the application of bike races. We assume that each sensor broadcasts messages which can only be received by sensors in a certain range and the radio transmission is under certain packet loss rates. As mentioned earlier, we can not define a priori routing paths for each sensor due to the high mobility of the bikers. Thus, the routing is basically performed by broadcasting and aggregation. In particular, the data collection procedure is divided into time slots of equal size within each of which sensor data are relayed to the sink by a multi-round routing algorithm. In each round in a time slot, each sensor aggregates messages heard in the previous round using CS and broadcasts the aggregated data.

More specifically, assume that there are $L$ rounds in each time slot. In the first round, sensor $i$ broadcasts a message of the format $(i,x_i)$ and receives a list of such messages from the 1-hop neighbors. In the following rounds, sensor $i$ computes and broadcasts a random combination of all received data from the previous round, given by 
\begin{equation}
 x^{l}_i = \sum_{j=1}^n a^l_{ij}x^{l-1}_j,\ l\geqslant 2,
 \label{csaggregation}
\end{equation}
where $x^1_i = x_i$ and $a^l_{ij}$ is a random Bernoulli coefficient and takes on the value of 1, -1 or 0. $a^l_{ij}=0$ means that no data from sensor $j$ is received by sensor $i$ in the previous round. The use of the Bernoulli random numbers allows the efficient coding of the coefficients with very few bits.

Furthermore, define $A_i^l=[a_{i1}^l,\ldots,a_{in}^l]$ which is a row vector, $A^l=[A_1^l;\ldots;A_n^l]$. Define $X^l=[x_1^l,\ldots,x_n^l]^t$. Thus, we have
\begin{equation}
 X^{l} = A^lX^{l-1}=A^lA^{l-1}X^{l-2}=\ldots=\prod_{r=2}^l A^rX^1,
 \label{csaggregation_matrixform}
\end{equation}
where $X^1=X$ and $X^l$ is in fact the aggregated sensor data in the $l$th round. Define $B^l=\prod_{r=2}^l A^r$, and we have $X^l=B^lX$. Let $B^l_i$ be the $i$th row of $B^l$, $B^l=[B^l_1;\ldots;B^l_n]$, and we have $x^l_i=B^l_iX$. For the consistency of the notation, we let $B^1$ be an identity matrix of $n\times n$.

Thus, we store at sensor $i$ $[A_i^2,\ldots,A_i^L]$ which consumes $n\times(L-1)$ bits. At the $l$th round, $l=1,\ldots,L$, sensor $i$ transmits a message of the form $(B^l_i,x^l_i)$. Note that both $B^l_i$ and $x^l_i$ can be computed recursively at sensor $i$ using received messages in the previous round by $B^l_i=\sum_j a_{ij}^l B_j^{l-1}$ and $x_i^l=\sum_ja_{ij}^lx_j^{l-1}$. As each entry of $B^l$, $b^l_{ij}$, is the sum of a number of products of a number of random Bernoulli numbers, $b^l_{ij}$ is actually a Binomial random number. If we control $L$ to be small and $A_i^l$ to be sparse, we can also encode $b^l_{ij}$ with few bits. For example, if we limit each message to aggregate at most $m$ data, $b^l_{ij}$ can be encoded by $\text{log}_2(m)$ bits. As a result, the length of the payload of the message in the $l$th round is
$$\text{bit}(B_i^l,x_i^l)=n\text{log}_2(m)+\text{bit(a real number)}.$$

\begin{figure}[t]
    \begin{center}
        \includegraphics[width=0.6\columnwidth]{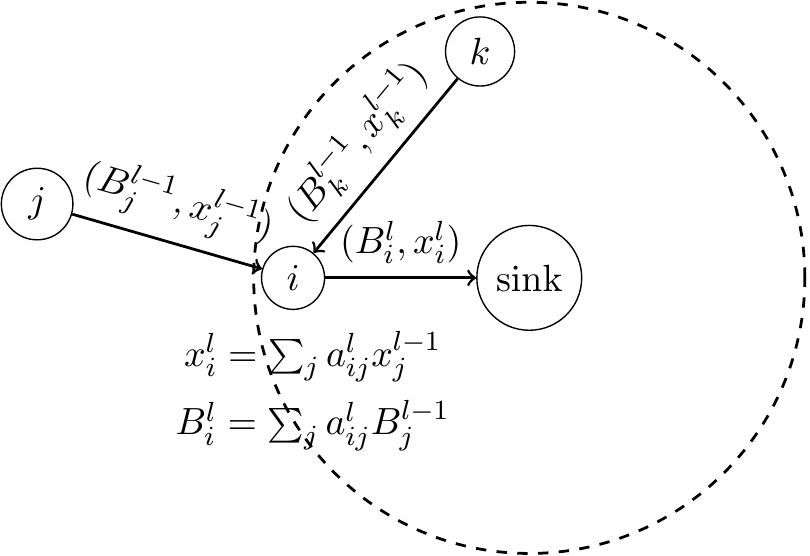}
    \end{center}
    \caption{Multi-round routing. Sensor $i$ stores $[A_i^2,\ldots,A_i^L]$ and broadcasts in the $l$th round a message of $(B_i^l,x_i^l)$ both of which are computed recursively by $B_i^l=\sum_{i}a_{ij}^lB_j^{l-1}$ and $x_i^l=\sum_ja_{ij}^lx_j^{l-1}$. Each message of $(B_i^l,x_i^l)$ received by the sink provides a linear equation $x_i^l=B_i^lX$ which is used to construct the linear system $Y=AX$. Messages from sensors out of the range of the sink such as $j$ are aggregated and relayed by other sensors.}
    \label{fig:csmultiroundrouting}
\end{figure}

The sink nodes receives messages in each round from sensors within the 1-hop range and every message of $(B^l_i,x^l_i)$ provides a linear equation $x^l_i=B^l_iX$ which is used to construct the linear system $Y=AX$. When the linear system is overdetermined, i.e. more than $n$ messages are received, we obtain $X$ by the analytical solution $X=(A^TA)^{-1}A^TY$. When it is underdetermined, we solve it with the CS formulation in eq. \ref{eq:cs_spatialsparsity} using a LP solver. Our multi-round routing algorithm is illustrated in Figure \ref{fig:csmultiroundrouting}. A limitation of such routing is that the $A$ in the resulted linear system $Y=AX$ is no more the random matrix required by the CS theory. However, we will show in the experiment section that empirically, we can still solve the system even under severe packet losses.

\section{Experiments}
 \label{sec:experiments}

\begin{figure}[t]
\centering
\subfigure[Biker motion, $x_i$, (change of location over time).]{%
 \includegraphics[width=0.45\textwidth]{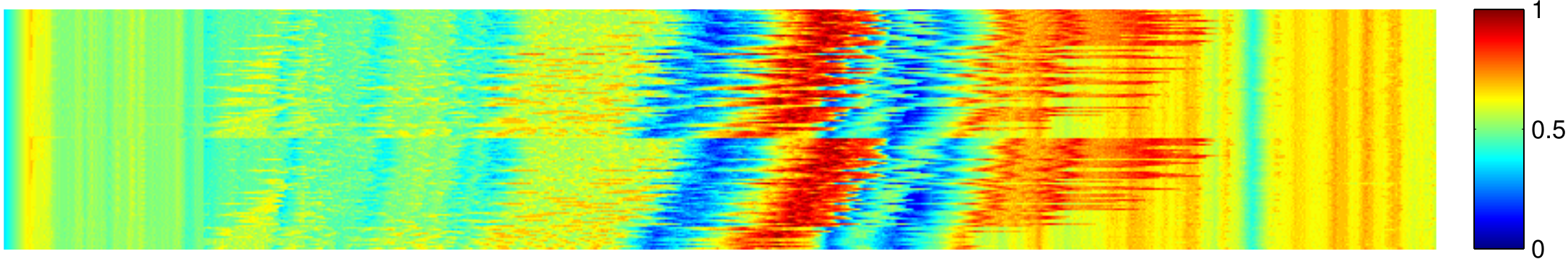}
\label{fig:bikermotion}}
\subfigure[Change of motion over time, $x_i(t)-x_i(t-1)$.]{%
 \includegraphics[width=0.45\textwidth]{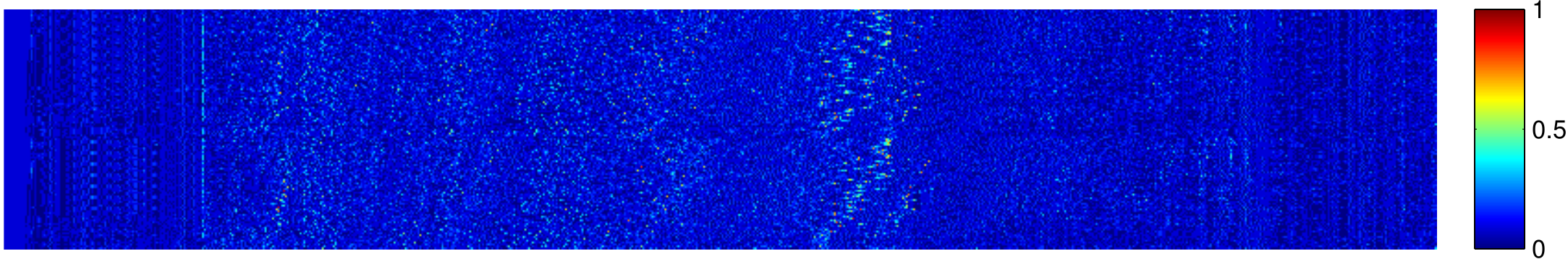}
\label{fig:motionchangeintime}}
\subfigure[Difference of motion between nearby bikers, $\sum_{ij\in E}||x_i-x_j||_1, \forall i$.]{%
 \includegraphics[width=0.45\textwidth]{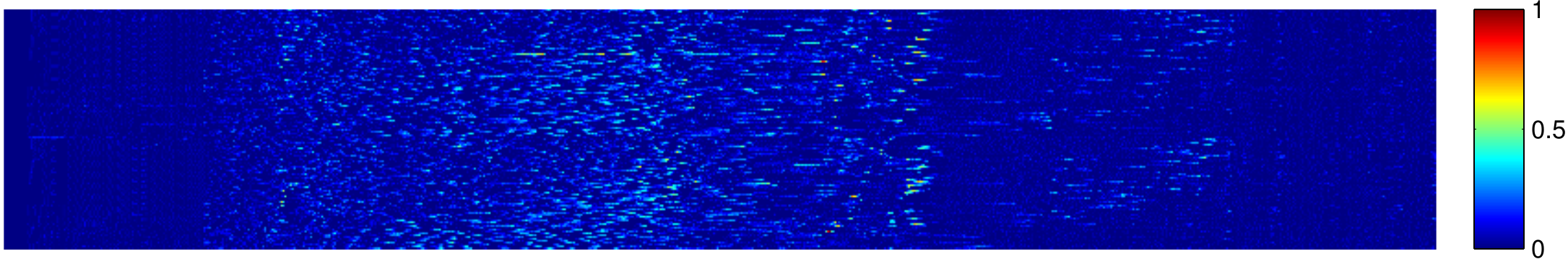}
\label{fig:motiondifferencebetweenneighbors}}
  \caption{Spatial and temporal sparsity of biker motion. The values are normalized in $\{0,1\}$, and the X axis is the time from 0 to 780 seconds and the Y axis is the biker/sensor ID indexed from 1 to $n=130$.}
  \label{fig:motionsparsity}
\end{figure}

We evaluated our approach on simulated data generated from real trajectories of motorbikes surrounding the peloton in the race of Paris-Tours 2013. In particular, we developed a bike race simulator based on a mobility model derived from the GPS recordings of the motorbikes. In the simulation, we control the speed of the group by the speed of the motorbikes and program individual behaviors of each biker by taking into account a set of parameters including fatigue level and objective. Thus, the simulator produces realistic trajectories of 130 bikers for a duration of 780 seconds, shown in Figure \ref{fig:motionsparsity}. Figure~\ref{fig:bikermotion} shows the motion (i.e. instant velocity) of the bikers calculated as the changes of locations over time, Figure \ref{fig:motionchangeintime} shows the changes of the motion of each biker over time, and Figure \ref{fig:motiondifferencebetweenneighbors} shows the differences between the motion of nearby bikers calculated as the total difference between each biker and its $10$ nearest neighbors in the space.

The performance or reconstruction accuracy is evaluated on the metric of stress, defined as
\begin{equation}
 stress(X,\hat{X})=\frac{\sum_i (x_i-\hat{x}_i)^2}{\sum_i x_i^2}.
 \label{eq:stress}
\end{equation}
The smaller the stress is, the better the reconstruction is.

\begin{figure}[t]
    \begin{center}
        \includegraphics[height=2.5cm, width=\columnwidth]{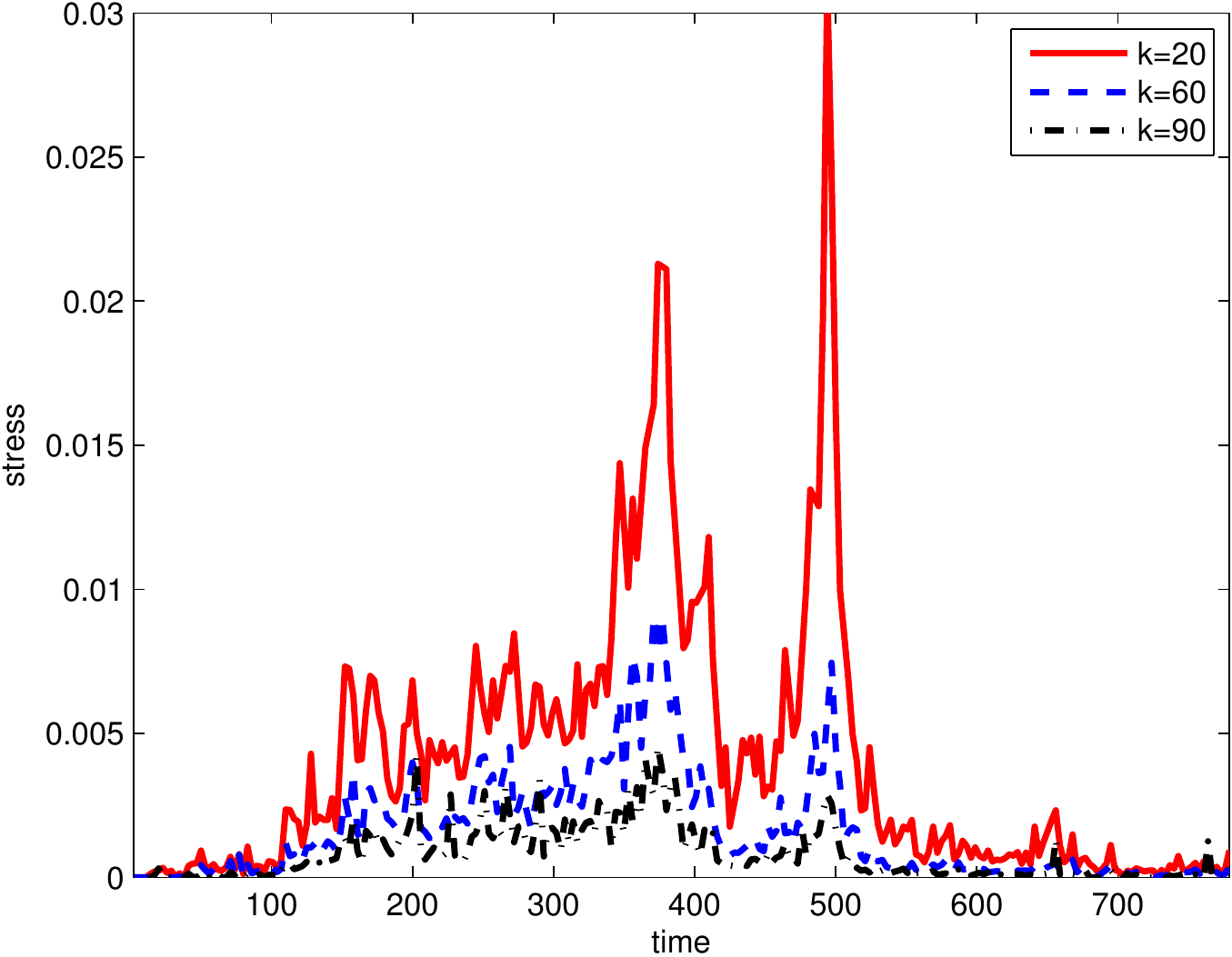}
    \end{center}
    \caption{Impact of $k$ on the accuracy of the reconstruction by using the CS method in eq. \ref{eq:cs_spatialsparsity} solved by a LP solver. We aggregate and reconstruct the data and compute the stress in each time of the simulation. The average stress in the entire simulation period is $0.4\%$, $0.16\%$ and $0.08\%$ for $k=20$, $60$ and $90$ respectively. For $k=60$, the reconstruction is the least accurate at time $374$, with a stress of $0.91\%$, when many bikers attack each other by accelerating irregularly, which breaks the assumption of the spatial sparsity.}
    \label{fig:accuracyonroutingtree}
\end{figure}

In the first experiment, we assume that the sink receives $Y=AX$, with $A$ being a random Bernoulli matrix of size $k\times n$ and its entries taking on either 1 or -1 with equal probability. This is the simplest case which is as good as collecting data via a routing tree on a static WSN with the architecture in Figure \ref{fig:csroutingtree}. We experiment with different $k$s in order to evaluate the impact of $k$ to the reconstruction accuracy, illustrated in Figure \ref{fig:accuracyonroutingtree}. It can be seen that the accuracy improves with the increase of $k$ which is intuitive. Overall, the reconstruction is fairly accurate for $k=60$ as the stress is reduced to less than $1\%$ which means that, for example, for the average velocity of 10 meter/second, the average error in the reconstruction is less than 1 meter/second.

\begin{figure}[t]
    \begin{center}
        \includegraphics[height=2.5cm, width=\columnwidth]{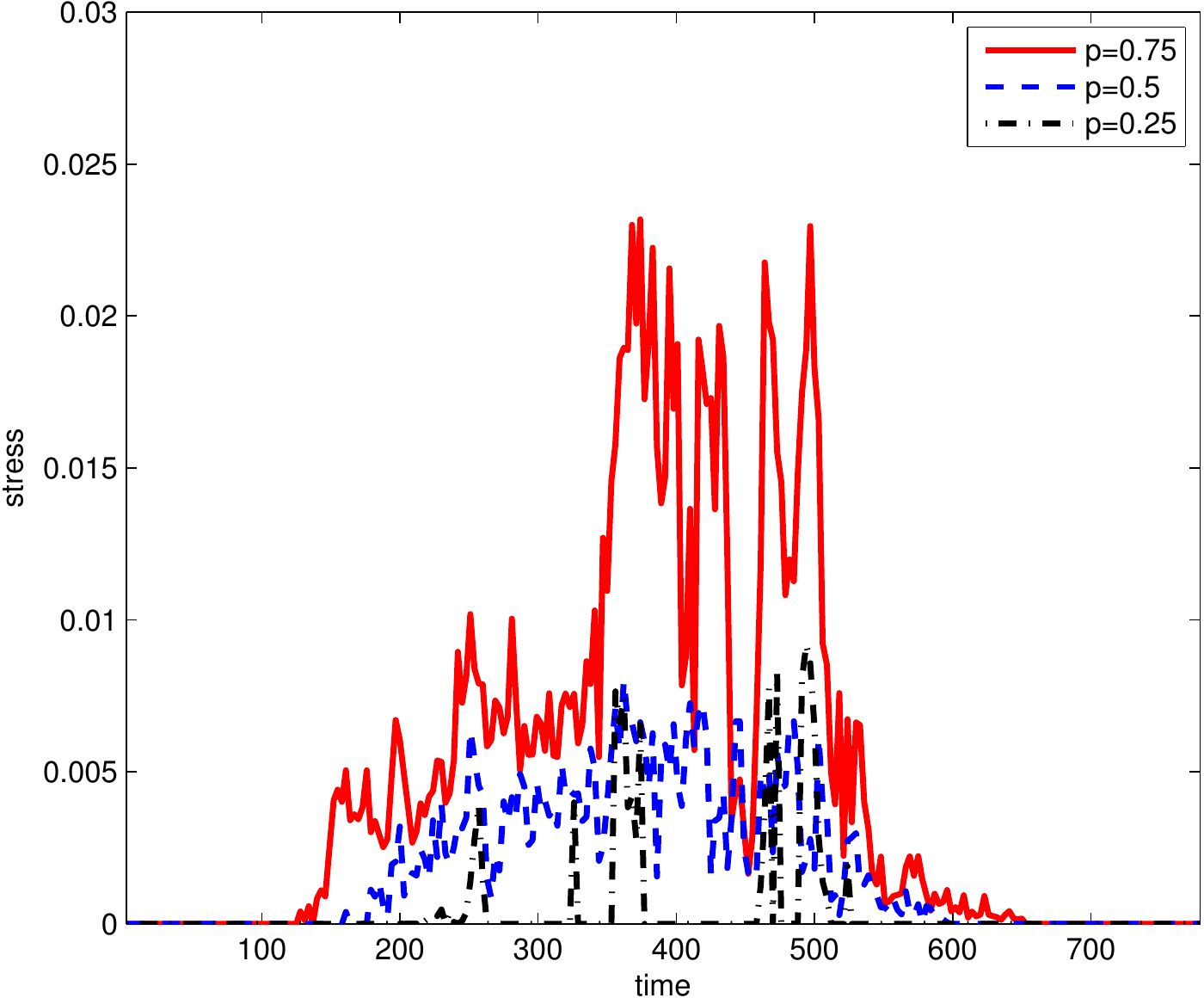}
    \end{center}
    \caption{Impact of packet loss rate $p$ on the accuracy of the reconstruction with the multi-round routing algorithm. The average stress in the entire simulation period is $0.48\%$, $0.14\%$ and $0.04\%$ for $p=0.25$, $0.5$ and $0.75$ respectively. For $p=0.5$, the reconstruction is the least accurate at time $365$, with a stress of $0.92\%$.}
    \label{fig:accuracyonmultiroundrouting}
\end{figure}

In the second experiment, we simulated the data collection procedure with the multi-round routing algorithm. In particular, we put two sink nodes in the front and at the back of the biker group respectively. We assume that the transmission range for each sensor is $50$ meters and that the number of rounds, $L$, in the multi-round routing algorithm is determined dynamically at each time so that every sensor can reach a sink node within $L$ rounds, $L\geqslant3$. While such control of $L$ is impossible in practice, in doing so, we make sure that each sensor data is aggregated and relayed to the sink with certain probability, which allows the evaluation of the reconstruction accuracy. In such a configuration, the experiment shows that, with the current location data, we always get overdetermined systems if there are no packet losses. Thus, we tune the packet loss rate for each connection so that each packet is dropped from a link randomly and independently with a probability $p$. We performed 10 simulations of the data routing for $p=0.25$, $0.5$ and $0.75$ respectively and found that even for $p=0.75$, i.e. on average $75\%$ packets are dropped, we still get overdetermined systems in about $30\%$ times due to the dense formation of the bikers in those times, and the underdetermined systems contain in the worst case about $20$ linear equations. Figure \ref{fig:accuracyonmultiroundrouting} shows the stresses over time under different $p$s. Overall, the reconstruction accuracy with the multi-round routing is comparable to that with the routing tree but looks more noisy. One important difference is that, while the multi-round routing often gathers more equations in $Y=AX$ even under severe packet loss rates, the combination matrix $A$ is not purely random as required by the CS theory which should have an effect on the accuracy. We are interested in a more thorough study on the impact of the routing.

\section{Conclusions and Future Work}
 \label{sec:conclusions}

This paper presents a CS-based approach to compressive data collection on mobile wireless sensor networks which is designed for the application of sensing in bike races. The method incorporates the problem-specific spatial sparsity and is integrated in a multi-round routing protocol based simply on broadcasting and aggregation. On motion data generated by a bike race simulator, our method achieved a fairly accurate reconstruction, with a stress of less than $1\%$, under a severe packet loss rate of $50\%$. Note that we did not compare our method with others simply because none of the previous work that we found are applicable in the situation of opportunistic routing in highly mobile and dynamic WSNs whereby each sensor has no knowledge of the next-hop sensors.

In the near future, we are interested in performing a test on real data from major bike races such as the Tour de France with which we intend to learn a better model of bike motions. Alternative to CS, networking coding is a technique which is also widely used for efficient data collection on WSNs. Thus, we also intend to investigate the feasibility of network coding in our application and compare it with CS-based approaches.

\bibliographystyle{IEEEtran.bst}
\bibliography{location}

\begin{thebibliography}{10}
\providecommand{\url}[1]{#1}
\csname url@samestyle\endcsname
\providecommand{\newblock}{\relax}
\providecommand{\bibinfo}[2]{#2}
\providecommand{\BIBentrySTDinterwordspacing}{\spaceskip=0pt\relax}
\providecommand{\BIBentryALTinterwordstretchfactor}{4}
\providecommand{\BIBentryALTinterwordspacing}{\spaceskip=\fontdimen2\font plus
\BIBentryALTinterwordstretchfactor\fontdimen3\font minus
  \fontdimen4\font\relax}
\providecommand{\BIBforeignlanguage}[2]{{%
\expandafter\ifx\csname l@#1\endcsname\relax
\typeout{** WARNING: IEEEtran.bst: No hyphenation pattern has been}%
\typeout{** loaded for the language `#1'. Using the pattern for}%
\typeout{** the default language instead.}%
\else
\language=\csname l@#1\endcsname
\fi
#2}}
\providecommand{\BIBdecl}{\relax}
\BIBdecl

\bibitem{tourdefrance_gps}
\emph{Dimension Data completes big data analytics and digital delivery platform
  for Tour de France}, Dimension Data, 2015, press release from
  \url{http://www.dimensiondata.com/}.

\bibitem{hikob}
\emph{Hikob}, \url{http://www.hikob.com/en/}.

\bibitem{cs}
D.~Donoho, ``Compressed sensing,'' \emph{Information Theory, IEEE Transactions
  on}, vol.~52, no.~4, pp. 1289--1306, 2006.

\bibitem{csintroduction}
E.~J. Cand\`{e} and M.~B. Wakin, ``{An Introduction To Compressive Sampling},''
  \emph{Signal Processing Magazine, IEEE}, vol.~25, no.~2, pp. 21--30, 2008.

\bibitem{compressivedatacollection}
C.~Luo, F.~Wu, J.~Sun, and C.~W. Chen, ``Compressive data gathering for
  large-scale wireless sensor networks,'' in \emph{Proceedings of the 15th
  Annual International Conference on Mobile Computing and Networking}, 2009,
  pp. 145--156.

\bibitem{dataaggregation}
L.~Xiang, J.~Luo, and C.~Rosenberg, ``Compressed data aggregation:
  Energy-efficient and high-fidelity data collection.'' \emph{IEEE/ACM
  Transactions on Networking}, vol.~21, no.~6, pp. 1722--1735, 2013.

\bibitem{cscluster}
R.~Xie and X.~Jia, ``Transmission-efficient clustering method for wireless
  sensor networks using compressive sensing,'' \emph{IEEE Trans. Parallel
  Distrib. Syst.}, vol.~25, no.~3, pp. 806--815, Mar. 2014.

\bibitem{mbcs}
R.~G. Baraniuk, V.~Cevher, M.~F. Duarte, and C.~Hegde, ``Model-based
  compressive sensing,'' \emph{IEEE Transactions in Information Theory},
  vol.~56, pp. 1982--2001, April 2010.

\bibitem{structuredsparsity}
F.~R. Bach, R.~Jenatton, J.~Mairal, and G.~Obozinski, ``Structured sparsity
  through convex optimization,'' \emph{CoRR}, vol. abs/1109.2397, 2011.

\bibitem{lasso}
R.~Tibshirani, ``{Regression shrinkage and selection via the lasso: a
  retrospective},'' \emph{Journal of the Royal Statistical Society: Series B
  (Statistical Methodology)}, vol.~73, no.~3, pp. 273--282, 2011.

\bibitem{flocking}
C.~W. Reynolds, ``Flocks, herds and schools: A distributed behavioral model,''
  in \emph{SIGGRAPH}, 1987, pp. 25--34.

\bibitem{graphsignal}
D.~I. Shuman, S.~K. Narang, P.~Frossard, A.~Ortega, and P.~Vandergheynst, ``The
  emerging field of signal processing on graphs: Extending high-dimensional
  data analysis to networks and other irregular domains,'' \emph{IEEE Signal
  Process. Mag.}, vol.~30, no.~3, pp. 83--98, 2013.

\end{thebibliography}

\end{document}